\documentstyle{article}%
\begin{document}
\title{Self-Organizing Approach for Finding Borders \\ 
of DNA Coding Regions}
\author{Fang Wu$^1$ and Wei-Mou Zheng$^2$\\
  {$^1$\it Department of Physics, Peking University, Beijing 100871, 
  China}\\
  {$^2$\it Institute of Theoretical Physics, Academia Sinica,
  Beijing 100080, China}}
\date{}
\maketitle

\begin{abstract}
A self-organizing approach is proposed for gene finding based on the
model of codon usage for coding regions and positional preference for
noncoding regions. The symmetry between the direct and reverse coding regions
is adopted for reducing the number of parameters. Without requiring prior
training, parameters are estimated by iteration. By employing the window
sliding technique and likelihood ratio, a very accurate segmentation is
obtained.
\end{abstract}

\leftline{PACS number(s): 87.15.Cc, 87.14.Gg, 87.10.+e}%

The data of raw DNA sequences is increasing at a phenomenal pace, providing 
a rich source of data to study. As a consequence, we now face the 
tremendous challenge of extracting information from the formidable volume 
of DNA sequence data. Computational methods for reliably detecting 
protein-coding regions are becoming more and more important. 

Genome annotation by statistical methods is based on various statistical
models of genomic sequences \cite{fick,fick2}, one of the most popular
being the inhomogeneous, three-period Markov chain model for 
protein-coding regions with an ordinary Markov model for noncoding 
regions. The independent random chain model can be included in this 
category by regarding it as a Markov chain of order 0. The codon usage model
is the independent random chain model of non-overlapping triplets, and
corresponds to an inhomogeneous Markov model of order 2. Signals
in a short segment are usually buried in large fluctuations. With well 
chosen parameters statistical models work as a noise filter to pick out 
the signals.

Methods based on local inhomogeneity, e.g. position asymmetry 
or periodicity of period 3, suffer fluctuations. 
Most of the current computer methods for locating genes require 
some prior knowledge of the sequence's statistical properties such as the 
codon usage or positional preference \cite{grant,fick3,karl}. That is, a
sizable training set is necessary for estimating good parameters of the
model in use \cite{boro1,boro2}.
Strongly biased by the training, such models have little power to discover
surprising or atypical features. Thus, it is desirable to decipher the 
genomic information in an objective way. Audic and Claverie \cite{audic} 
have proposed a method which does not require learning of species-specific
features from an arbitrary training set for predicting protein-coding
regions. They use an {\it ab initio} iterative Markov modeling procedure
to automatically partition genome sequences into direct coding, reverse 
coding, and noncoding segments. This is an expectation-maximization (EM) 
algorithm, which is useful in modeling with hidden variables, and is
performed in two steps of expectation and maximization \cite{baldi,law,car}.
Such a self-organizing or adaptive approach uses all the available
unannotated genomic data for its calibration. 

Before introducing the model we use and describing the technical details, we
explain the EM algorithm with a simple pedagogic model which assumes
that a DNA sequence written in four letters $\{a, c, g, t\}$ is generated by 
independent tosses of two four-sided dice. An 
annotation maps the DNA sequence site-to-site to a two-letter sequence of 
the alphabet $\{C, N\}$ ($C$ for coding and $N$ for noncoding). Two sets 
$\{p_a, p_c, p_g, p_t\}$ and $\{q_a, q_c, q_g, q_t\}$ of positional nucleotide 
probabilities are associated with the two dice $C$ and $N$, respectively.
The total probability for the given DNA sequence $S=s_1s_2\ldots$ to be seen 
under the model is the partition or likelihood function
\begin{equation}
Z=\sum_H P(S|H_\alpha )=\sum_H \prod_i P(s_i|h_i^\alpha ),
\end{equation}
where the summation is over all the possible ``annotations" 
$H=\{H_\alpha \}$ with $H_\alpha =h_1^\alpha h_2^\alpha \ldots$, $h_i^\alpha 
\in\{N,C\}$, and $P(s|C)=p_s$, $P(s|N)=q_s$. The unknown two sets of 
probabilities can be determined by maximizing the likelihood $Z$. From 
Bayesian statistics
\begin{equation}
P(H_\alpha |S)=\frac {P(S|H_\alpha )P(H_\alpha )} {\sum_H P(S|H_\alpha )
P(H_\alpha )},
\end{equation}
with prior $P(H_\alpha )$ assumed, the most possible $H_\alpha$ can then 
be selected as the inferred annotation. As we know, coding regions are 
organized in blocks. The first simplification is the window 
coarse-graining. The sequence $S$ is divided into nonoverlapping window 
segments of constant length $w$, and each whole window is entirely assigned 
to either $N$ or $C$. Conducting Bayesian analysis for window $W_j$ 
and accepting uniform prior, we have
$P(h|W_j) \propto P(W_j|h) $. 
The second simplification is to introduce ``temperature" $\tau$ (as in 
the simulated annealing), 
replace $P(W_j|h_j)$ with $[P(W_j|h_j)]^{1/\tau}$ and take the limit 
$\tau\to 0$. In this way we keep only a single term, i.e. the greatest one, 
in the summation for $Z$. Window $W$ is inferred to belong to either $N$ 
or $C$ depending on whether $P(W|N)$ or $P(W|C)$ is larger. The likelihood 
maximization is then equivalent to estimating nucleotide probabilities with 
frequencies in two window classes inferred from the pre-assumed $\{p_s\}$ and 
$\{q_s\}$. Consistency requires 
that the estimate probabilities must be equal to $\{p_s\}$ and $\{q_s\}$. 
This ``fixed point" can be found by iteration. As an example, we use the 
first $99\times 5\,051 = 500\,049$ nucleotides of the complete genome
of E.~coli as the input data. Statistical significance requires
that the window size cannot be too small, while a large window size would give 
poor resolution in discriminating different regions. The window size is 
chosen to be $w=99$. We assign the $5\,051$ fixed nonoverlapping windows 
to the two subsets of $N$ and $C$ in either a periodic or a random way. 
We estimate $p_s$ and $q_s$ from the counts of different nucleotides 
in each subset. The likelihood functions for
each window are then calculated using the estimated $p_s$ and 
$q_s$, and the assignment of the windows to $C$ or $N$ is updated 
according to which of $P(W|C)$ and $P(W|N)$ is larger. This ends one 
iteration. The process of iteration converges to a single fixed point of
precision $10^{-4}$ around step 28 for different initializations with a final
window assignment to $N$ and $C$ also given.
The final $p_s$ and $q_s$ are $\{0.219, 0.270, 0.289, 
0.222\}$ and $\{0.279, 0.213, 0.227, 0.281\}$. The $q_s$ estimated 
from the complete genome are $\{ 0.285, 0.214, 0.218, 0.283\}$, which are 
rather close to the corresponding convergent values. 

More realistic models take the three phases in the coding regions and the 
opposite ordering of the direct and reverse coding regions into 
account. Such models adopt 7 subsets: one for noncoding (N), three for 
direct coding (C$_1$, C$_2$, C$_3$) and three for reverse coding (C$_4$, 
C$_5$, C$_6$). The subscript $i$ in C$_i$ indicates the phase 0, 1 or 2 
accordng to $i$ (mod 3). From the genomic data statistics, we may assume
that there is symmetry between the direct and reverse coding regions, which
means that a reverse coding sequence is indistinguishable from a direct 
coding sequence if we make the exchanges $a\leftrightarrow t$, 
$c\leftrightarrow g$ and reverse the order. For the model based on the 
positional preference of codons, instead of 7 sets of positional 
nucleotide probabilities, we need only 4 sets. The reduction of the total
number of parameters by the symmetry consideration improves
the statistics. The procedure of
iteration is similar to that for the last model, the only difference being 
that now we have to estimate 4 sets of probabilities and calculate 7 
likelihood functions for a window. 

We use a better model based on the codon usage. We now need a set of 
64 probabilities for coding regions. For noncoding segments, 4
positional nucleotide probabilities are used just as before. To simplify
the programming, we move the windows with a phase-shift other than zero by one 
or two nucleotides to clear the phase-shift, although we can calculate the 
marginal distribution probabilities for uni- and bi-nucleotides. For example,
we replace the window $W=s_is_{i+1}\ldots s_{i+w-1}$ marked as C$_2$ with
$W'=s_{i+1}s_{i+2}\ldots s_{i+w}$. (The
alternative way is to consider a cyclic transformation.) Our further 
discussions are all based on this model. It is observed that the iteration 
also quickly converges to a fixed point. Contrary to the two-sets model where 
coding and noncoding are symmetric, and extra knowledge is required to 
relate one set to coding and the other to noncoding, we can now distinctly 
distinguish coding from noncoding regions, even with their phases fixed. 
Direct and reverse coding sets are symmetric in the model. 
However, the fact that stop codons $taa$, $tag$ and $tga$ are rare can 
be used to remove the symmetry between direct and reverse coding. That 
is, if the convergent probabilities for $taa$, $tag$ and $tga$ are all 
significantly small in comparison with the other 61, sets C$_1$, C$_2$ and
C$_3$ then do indeed correspond to direct coding. (Otherwise, those of $tta$, 
$cta$ and $tca$ would be small instead.) 

We employ the sliding window technique to improve the resolution as follows. 
We shift each window by 3 nucleotides, initiate the window assignment with 
the convergent probabilities just obtained, and then find new assignments 
for the shifted windows by iteration. We repeat the shifting process 32 
times to cover the window width. This ends with 33 assignments for 
triplets, except for a few sites at the two ends. By a majority vote we 
can obtain a triplet assignment of the whole sequence. 

Recently, an entropic segmentation method that uses the Jensen-Shannon
measure for sequences of a 12-letter alphabet has been proposed to find
borders between coding and noncoding regions \cite{stan}. Their best 
result was obtained on the genome of the bacterium {\it Rickettsia 
prowazekii}. We test our approach with the same genome data. 
To inspect the accuracy we obtain the ``true" 
assignment of sites based on the known annotation as follows. If a
nucleotide is in a noncoding region, it belongs to N. If it is in a coding
(or reverse coding) region and the site-index of the beginning nucleotide
plus 1 is congruent to $i$ modulo 3, the nucleotide under consideration
will belong to C$_{1+i}$ (or C$_{4+i}$). For overlapping coding zones we 
may keep two alternative assignments. We define three rates of accuracy 
$R_2$, $R_3$ and $R_7$: $R_2$ only discriminates coding from noncoding 
segments while $R_7$ covers full discrimination of the 7 sets, and $R_3$ 
ignores the phases. For the total $N =1\,111\,523$ nucleotides, we obtain
$R_2=91.7\%$, $R_3=89.8\%$ and $R_7=89.7\%$. (The rates without window sliding
are $R_2=89.1\%$, $R_3=84.8\%$ and $R_7=84.4\%$.)

For finding block borders, to eliminate illusary fluctuations  
we accept only the assignments with the 33 identical samplings, and 
regard others as undetermined. When two adjacent identified blocks are of 
the same assignment we join the two together with the sites between into a 
single zone of the same assignment. Otherwise, we take the middle site of 
the intervening undetermined zone as the border, and assign the two sides
according to their corresponding flank blocks. We can do
the job better by means of the likelihood ratio. Suppose that the left 
block is assigned to $l$, and the right to $r$. A point $m$ in the 
intervening zone divides the zone into two segments $L_m$ and $R_m$. The 
likelihood ratio is defined as 
\begin{equation}
\Gamma_m =\frac {P(L_m|l)P(R_m|r)}{P(L_m|r)P(R_m|l)}.
\end{equation}
The maximal $\Gamma_m$ places the border at $m$. This segmentation 
finally gives the accuracy rates $R_2=93.3\%$, $R_3=92.8\%$ and $R_7=92.7\%$. 

In Ref.~\cite{stan} the quantity quantifying the coincidence between 
borders inferred from the segmentation and those from the known annotation is 
defined by
\begin{equation}
D=\frac 1{2N}\left[ \sum_i \min_j|b_i-c_j|+\sum_j \min_i|b_i-c_j| 
\right],
\end{equation}
where $\{b_i\}$ is the set of all borders between coding and noncoding 
regions, and $\{c_j\}$ is the set of all cuts produced by the 
segmentation. We use an even harsher quantity $D$ by interpreting $\{b_i\}$ 
and $\{c_j\}$ as the borders of all coding zones. That is, we include 
borders of each overlapping coding zone. The total number of ``CDS" in 
the annotation is 834, one of which has two joint zones. We obtain  
$1-D=87.7\%$, compared with $\sim 80\%$ of Ref.~\cite{stan}. In Fig.~1 we show
a comparison of the inferred segmantation with the known coding regions.
In the section from $475\,500$ to $497\,500$ there are two overlaps (one
for direct, and the other for reverse coding regions), and the shortest gap
separating adjacent coding regions is just 1 nucleotide (at $486\,215$). They
do not escape detection. As mentioned in \cite{stan}, there are
two very close coding regions in the same phase ($538\,197:539\,879$ and
$539\,937:540\,887$). The result from the majority vote is shown in Fig.~2 for
the section. We see indeed a peak of the counts for set N between the two coding
regions. The highest count for N is 32, and so is ignored in our strategy.
There is indeed plenty of room for improving this approach. A larger width $w=123$ 
gives higher accuracy rates: $R_2=93.6\%$, $R_3=93.3\%$, $R_7=93.0\%$ and 
$1-D= 88.2\%$. When we consider only the triplets with all 33 assignments
identical in window sliding the rates are $R_2=98.7\%$, $R_3=98.6\%$ and
$R_7=98.6\%$. In the above we avoid setting up an arbitary cut-off threshold.
If a threshold of 17 counts is used to determine the segments whose central
parts have 33 identical samplings,
for $w=99$ we predict a total of $1\,001\,351\ (90.1\%)$ sites with
accuracies $R_2= 97.4\%$ and $R_3= 95.4\%$. The accuracy rate for noncoding
regions is 96.5\%, much
higher than that of Ref.~\cite{audic}. It is important and feasible to
integrate biological signals into our algorithm. We expect our algorithm,
with certain modifications, should work well for other species, too.

\begin{quotation}
{ This work was supported in part by the Special Funds for Major National  
Basic Research Projects, the National Natural Science Foundation 
of China and Research Project 248 of Beijing.}
\end{quotation}


\begin{figure}[hb]
\caption{
Comparison between the inferred segmentation (dotted lines) and the known
coding regions of {\it Rickettsia} (shaded areas).}
\end{figure}

\begin{figure}[ht]
\caption{
Counts of majority assignment in the section containing two very close
coding regions (shaded areas) in the same phase. A peak corresponding to
noncoding assignment is clearly seen.} 
\end{figure}

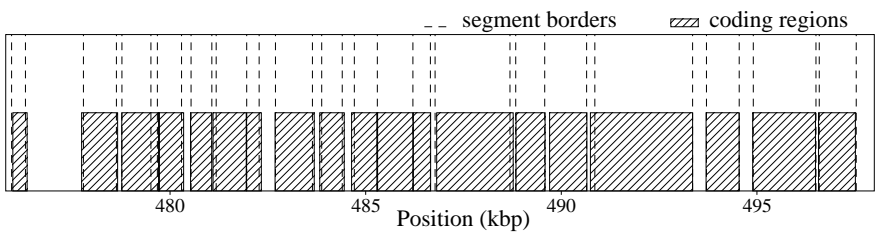
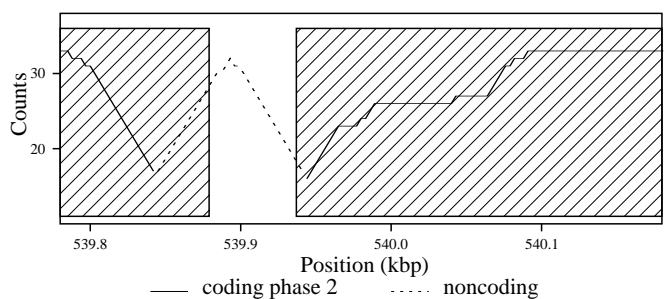
\end{document}